\documentclass{PoS}

\title{Chiral Quark Dynamics and the Ramond-Ramond $U(1)$ Gauge Field}

\ShortTitle{Chiral Quark Dynamics and the Ramond-Ramond $U(1)$ Gauge Field}

\author{\speaker{H. B. Thacker}%
         \thanks{}\\
        University of Virginia\\
        E-mail: \email{hbt8r@virginia.edu}}

\abstract{Topological excitations in the QCD vacuum take the form of coherent codimension one sheets with positive
and negative sheets juxtaposed in a dipole layer. These sheets may be interpreted as Luscher's ``Wilson bags,'' which 
are domain walls between discrete quasivacua labelled by a local value of the $\theta$ parameter equal to $2\pi k$, with $k$ 
given by the number of units of background Ramond-Ramond flux. This picture of the vacuum is closely analogous to Coleman's description 
of the 2D massive Schwinger model, where $\theta$ is interpreted as a background electric field, and a pointlike charged particle 
is a domain wall between vacua which differ by one unit of background electric flux. The main effect of the Ramond-Ramond $U(1)$ field
in low energy QCD dynamics is the generation of phenomenologically important contact terms: (1) The dominant contact term in the 
topological charge correlator which leads to positive topological susceptibility, (2) The $\eta'$ mass insertion, and (3) An
$SU(N_f)\times SU(N_f)$ invariant Nambu-Jona Lasinio 4-quark interaction. 
         }

\FullConference{XXIX International Symposium on Lattice Field Theory \\
		 July 10 ~V 16 2011\\
		 Squaw Valley, Lake Tahoe, California}

\begin{document}

\section{Introduction}

The dynamical mechanism which spontaneously breaks chiral symmetry and forms the quark condensate in QCD is still 
poorly understood. The Nambu-Jona Lasinio (NJL) model is a phenomenologically successful model of 
chiral symmetry breaking, but it is not clear how the local 4-fermion interaction which drives chiral symmetry breaking in the NJL model arises
from the gauge interactions of QCD. It can be argued that the topological charge excitations of
the color gauge field play a central role in forming the chiral condensate. According to the Banks-Casher relation, there will be a quark condensate with 
$\langle \bar{q}q\rangle\neq 0$ only if the density $\rho(\lambda)$ of Dirac eigenmodes in typical gauge fields remains constant for $\lambda\rightarrow 0$, 
i.e. there must be a large density of ``near zero'' eigenmodes.  For a localized instanton, the associated 'tHooft zero mode can be 
thought of as a quark bound to the instanton, with left-handed and right-handed quarks
attracted by instantons and anti-instantons respectively. A gas of instantons and anti-instantons would produce a band of near-zero eigenmodes and thereby
a quark condensate \cite{Diakanov84}.  But this idea is more general than the instanton model. The near-zero Dirac eigenmodes that form the condensate should arise
from whatever gauge fluctuations contribute to the topological susceptibility of the QCD
vacuum. Monte Carlo studies of both 4D SU(3) Yang-Mills theory \cite{Horvath03} and 2D $CP^{N-1}$ sigma models \cite{Ahmad05,Keith-Hynes08} 
have indicated that the topological charge distribution in typical gauge configurations is in fact dominated by thin codimension one
sheets of topological charge, with positive and negative sheets juxtaposed in a dipole layer. The thickness of the sheets
goes to zero in the continuum limit, while the confinement scale seems to be set by the distance over which a normal vector to the sheets decorrelates,
due to the twisting and bending of the sheets.
These topological charge sheets provide a natural mechanism to form the chiral condensate from the $\lambda\approx 0$ surface quark eigenmodes on the sheets.

In this talk, I will discuss some of the implications for chiral dynamics which follow from the interpretation of these sheets as D6 branes in 
holographic QCD. D6 branes are magnetic sources of the Ramond-Ramond U(1)
gauge field in IIA string theory. In holographic QCD, a D6 brane induces an excitation of the Chern-Simons tensor of the color gauge field on the 3-dimensional 
intersection of the D6 brane with the D4 color branes. The interaction between the D6 brane and the color Chern-Simons tensor is dictated by
anomaly inflow arguments \cite{Green96}.
I will show that the physical, low energy effect of the D6 branes and the associated Ramond-Ramond gauge interaction in QCD is to produce certain phenomenologically important contact terms:
(1) The dominant contact term in the topological charge correlator that is required in order to have $\chi_t>0$, and (2) A 4-quark contact term that is flavor
singlet in the RR exchange channel (see Fig. 1). For s-channel exchange ($q\bar{q}$ annihilation) this contact term produces the $\eta'$ mass insertion,
while t-channel RR exchange provides a $U(N_f)\times U(N_f)$ symmetric NJL interaction. All these contact terms can be said to arise from singular gauge configurations.
Note that in the field theoretic limit of the holographic model,
the RR field does not represent any new degree of freedom beyond the Yang-Mills fields. It is rather an auxiliary field that 
couples to the Chern-Simons tensor and desribes the singular, sheet-like fluctuations of the gauge fields in the QCD vacuum. 
It plays a role similar to that of the auxiliary massless Goldstone field that cancels the massless Chern-Simons pole in 
a covariant treatment of the 2D Schwinger model (the ``Kogut-Susskind dipole'' cancellation)\cite{Kogut75}. 

Long ago it was pointed out by Witten that the instanton model is incompatible with large-$N_c$ chiral dynamics \cite{Witten79a}, since instantons
are exponentially suppressed at large $N_c$, while consistency of the large $N_c$ expansion requires that the
$\eta'$ mass squared be of order $1/N_c$. Moreover, the form of the axial U(1) anomaly term in the chiral Lagrangian generated by instantons
would be $\propto$ Det $U$ (where $U$ is the $U(N_f)\times U(N_f)$ chiral field), while phenomenology indicates that it is of the form ($\log$ Det $U)^2$,
i.e. a pure $\eta'$ mass term. The multivaluedness of the $\log$ in this expression is an indication of the existence of discrete, quasistable ``k-vacua''
in which the effective local $\theta$ parameter is shifted by integer multiples of $2\pi$ \cite{Witten_largeN}.
In the large-$N_c$ limit, the vacua with $\theta = 2\pi k$ for $k\neq 0$ are long-lived and nearly degenerate with the $k=0$ vacuum. In Witten's analysis, topological
fluctuations in the large-$N_c$ gauge theory vacuum appear not as localized instantons, but rather as domains of $k\neq 0$ vacua separated by codimension one
domain walls. 

After the advent of holographic QCD, Witten demonstrated that these domain
walls are related to wrapped D6 branes in type IIA string theory \cite{Witten98}. In the Witten-Sakai-Sugimoto \cite{Witten98,Sakai04} formulation of HQCD, the color gauge fields are the open string
gauge fields on $N_c$ coincident D4 branes (5-dimensional objects) on $R_4\times S_1$. Here $R_4=$ spacetime, and the $S_1$ is the compactified 5th dimension of the D4 branes.
A domain wall in the 4D color gauge theory corresponds holographically to a D6 brane wrapped around the $S_4$
and intersecting with the color D4 branes in a 3-dimensional surface in spacetime (an ``I2 brane''). 

These theoretical arguments for codimension one topological charge excitations have been strongly supported by lattice studies of topological charge distributions \cite{Horvath03,Ilgenfritz}.
These studies showed that the topological charge density in pure glue SU(3) gauge theory exhibits a laminated structure consisting of juxtaposed, alternating sign sheets
of topological charge. These sheets lead to a characteristic structure of the two-point topological charge correlator which is dominated by a positive contact term and a 
short range negative tail. The topological susceptibility $\chi_t$ is given by the integrated correlator. Spectral positivity arguments require that the correlator be {\it negative}
for any finite separation $x\neq 0$. This means that the correlator must have a dominant positive contact term at $x=0$ in order to have positive $\chi_t$. 
The short range negative tail arises because of the juxtaposition of positive and negative sheets. (The Chern-Simons excitation induced by the D6 brane is a dipole layer
of topological charge.) 
In the Monte Carlo studies, the contribution to $\chi_t$ of the positive contact term and that of the negative tail are separately divergent in the continuum limit, but there is a delicate cancellation
between them, leaving a $\chi_t$ which is positive and scales properly in the continuum limit \cite{Horvath05,Lian07}. 
The regular laminated arrangement of alternating sign sheets that is reponsible for the peculiar structure of the TC correlator resembles a phenomenon that arises in string theory known as a
``tachyonic crystal'' \cite{Polchinski94,Callan94,Thacker10}. 

\section{Domain walls in QCD and Ramond-Ramond charge in string theory}

As I will discuss in this talk, the domain walls suggested by Witten in 4D gauge theory consists of sheet-like
excitations of the Chern-Simons tensor,
\begin{equation}
\label{eq:CStensor}
{\cal K}_{\mu\nu\lambda} = {\rm Tr} \left(A_{\mu}F_{\nu\lambda}-\frac{1}{3}A_{\mu}A_{\nu}A_{\lambda}\right)
\end{equation}
In a remarkable early paper, Luscher \cite{Luscher78} emphasized the importance in 4D Yang-Mills theory of the integral of the Chern-Simons tensor over a 3-dimensional surface $\sigma$ in 4D spacetime,
\begin{equation}
\label{eq:Wbag}
B(\sigma) = \exp\left(\frac{e}{\pi}\int d\sigma_{\mu\nu\lambda}{\cal K}^{\mu\nu\lambda}\right)
\end{equation}
Luscher argued that this integral should be interpreted as the world volume of a (2+1)-dimensional membrane or ``Wilson bag.'' This construction clarifies a very 
instructive analogy between topological charge structure in 4D Yang-Mills theory and that of 2-dimensional U(1) gauge theories such as the massive Schwinger model and the $CP^{N-1}$ sigma models. 
The Wilson bag integral (\ref{eq:Wbag}) is the 4D Yang-Mills analog of an ordinary Wilson line in the 2D U(1) theory. In the same sense that the Wilson line can be intepreted
as the gauge phase associated with the world line of a pointlike charged particle, the Wilson bag integral should be seen as a phase attached to the world volume of a 2-dimensional
membrane. 

Before the role of topology in 4D Yang-Mills theory was understood, the significance of the $\theta$ parameter in gauge theories with topological charge was first clarified 
by Coleman in the 2D massive Schwinger model \cite{Coleman76}. In this model, $\theta$
can be interpreted as a background electric field. With this interpretation, the 2D version of domain walls between k-vacua is easily understood. A pointlike charged particle in one 
space dimension is essentially a domain wall between vacua which differ by one unit of electric flux. The k-vacua are vacua containing k units of electric flux, and the requirement
that $\theta$ jumps by $\pm 2\pi$ across the domain wall is just Gauss's law. Note that in 2D, one-photon exchange produces a linear potential which confines charge. If a $q\bar{q}$ pair 
appears in the k=0 vacuum, it has a unit string of electric flux between the quark and antiquark, i.e. a bubble of $k=\pm1$ vacuum. The decay of a $k\neq 0$ vacuum is just string breaking, i.e.
screening by the production of quark pairs. For finite quark mass the string is quasistable, and virtual quark pairs in the vacuum will produce regions of nonzero topological
charge (= electric field in the 2D case) and hence $\chi_t>0$. As $m\rightarrow 0$, screening of $k\neq 0$ vacua is complete, and the topological susceptibility goes to zero. 

Our discussion of topological fluctuations in large-$N_c$ QCD is a direct 4-dimensional generalization of the Schwinger model analysis. 
The holographic description of QCD makes this analogy complete by identifying the charged domain walls (analogous to electrically charged particles of the Schwinger model)
as wrapped D6 branes. The famous result of Polchinski \cite{Polchinski95} which started the Dbrane revolution in string theory was a demonstration that D6
branes are physical entities that carry Ramond-Ramond charge in IIA string theory. They are sources of the U(1) gauge field associated with the massless state
of a closed Ramond-Ramond string. The Wilson bag integral (\ref{eq:Wbag}) plays a role in 4D QCD 
analogous to a Wilson loop in the Schwinger model. Following this analogy, we find that the $\theta$ parameter in 4D QCD can be interpreted as a background
Ramond-Ramond (RR) gauge field \cite{Witten98} and that it plays a role similar to the background electric field in Coleman's anaysis of the Schwinger model. 

The relation between k-vacua and the Wilson bag integral (\ref{eq:Wbag}) is easily seen. The topological charge is the divergence of the Chern-Simons current,
i.e. $Q=\partial^{\mu}K_{\mu}$, where 
\begin{equation}
K_{\mu}=\frac{1}{8\pi^2}\epsilon_{\mu\nu\lambda\sigma}{\cal K}^{\nu\lambda\sigma}
\end{equation}
(or $K_{\mu}=\frac{1}{2\pi}\epsilon_{\mu\nu}A^{\nu}$ in the 2D case).
If we add a $\theta$ term to the action where $\theta$ is a nonzero constant over a closed finite region $R$ with boundary $\partial R$ and zero outside $R$, this is equivalent
to adding a Wilson bag term integrated over the boundary,
\begin{equation}
S_{\theta} = \theta\int_R d^4x\; Q(x) = \theta\int_{\partial R}d\sigma_{\mu\nu\lambda}{\cal K}^{\mu\nu\lambda}
\end{equation}
For the 2D model, the right-hand side is a Wilson loop around the boundary,
\begin{equation}
S_{\theta} = \theta\oint_{\partial R} dx_{\mu} A^{\mu}
\end{equation}
Thus for both the 2D and 4D theories, a closed Wilson bag operator with unit charge creates a bubble of $\theta=\pm 2\pi$ vacuum inside the bag. 

\section{Ramond-Ramond exchange and contact terms in QCD}

From the fact that topological charge is the divergence of the Chern-Simons current, it is easy to show that, in order to have nonzero topological susceptibility $\chi_t$,
the Chern-Simons current correlator must have a massless pole at $q^2=0$,
\begin{equation}
\label{eq:CScorr}
G_{\mu\nu}(q) \equiv \int d^4x e^{iq\cdot x}\langle K_{\mu}(x)K_{\nu}(0)\rangle \stackrel{q\rightarrow 0}{\sim} -\chi_t\frac{q_{\mu}q_{\nu}}{q^2}
\end{equation}
In the Schwinger model, this pole arises from one-photon exchange. The gauge invariant axial current $\hat{j}_5^{\mu}$ is a combination of the conserved
fermion current and the Chern-Simons current $K_{\mu}$. 
\begin{equation}
\label{eq:KSdipole}
\hat{j}_5^{\mu} = j_5^{\mu} + K^{\mu} 
\end{equation}
In matrix elements of the gauge invariant axial current, the Chern-Simons pole (\ref{eq:CScorr}) 
is cancelled by an unphysical massless Goldstone pole
which couples to the conserved current $j_5^{\mu}$.  
(The introduction of auxiliary massless fields 
may be avoided by quantizing in non-covariant Coulomb gauge, but at the price of having a nonlocal action with a linear potential between quarks.)
Eq. (\ref{eq:KSdipole}) expresses the fact that a gauge invariant description of the quark must include both the bare fermion current and the comoving gauge excitation,
represented by the CS current. Either of these excitations has a massless pole in its correlator, but the poles cancel for the gauge invariant
combination (\ref{eq:KSdipole}). 

\begin{figure}
\vspace*{2.50cm}
\includegraphics{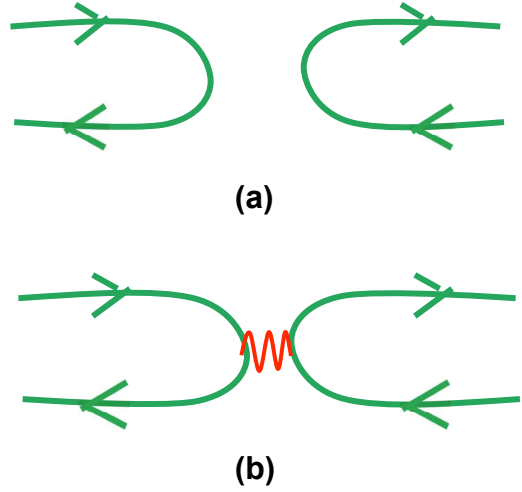}
\vspace{4.5cm}
\caption{(a) The quenched double hairpin correlator which measures the gluonic $\eta'$ mass insertion.
The indicated quark propagators are assumed to be summed over all gauge field configurations.
(b) The s-channel RR exchange picture for the hairpin correlator. RR exchange results in a local
4-quark contact interaction due to its derivative coupling to the chiral field. These $q_{\mu}$ factors
cancel the massless pole and convert it to a delta-function.}
\label{fig:hairpin}
\end{figure}
 
In QCD, the CS correlator also must have a massless pole, since $\chi_t\neq 0$. As in the Schwinger model, this pole does not appear in gauge
invariant amplitudes. Nevertheless, it has direct physical implications: it produces the gauge invariant contact term in the topological
charge correlator,
\begin{equation}
\label{eq:chi_t}
\chi_t = \int d^4x\langle Q(x) Q(0)\rangle = -\lim_{q\rightarrow 0} q^{\mu}q^{\nu}G_{\mu\nu}(q)
\end{equation}
By introducing quarks as probes of the gauge field (in the quenched approximation), we can identify the RR field $\theta(x)$ with the U(1) chiral
field $\eta'/f_{\pi}$. Similarly, the axial vector current $j_5^{\mu}$ is identified with $\partial_{\mu}\theta$, and hence with 
the RR field strength $f$ integrated around the compact direction $x^4$,
\begin{equation}
\partial_{\mu}\theta = \partial_{\mu}\oint_{S_1}a_4\; dx^4 = \oint_{S_1}f_{\mu4}dx^4
\end{equation}
There is a Kogut-Susskind dipole cancellation between the $\partial_{\mu}\theta$ field which describes a domain wall, and the Chern-Simons current $K_{\mu}$ describing
the comoving Yang-Mills excitation. The sum of the two gives the $\mu,4$ component of the gauge invariant RR field strength,
\begin{equation}
\label{eq:RRfieldstrength}
\oint_{S_1}\hat{f}_{\mu 4}= \partial_{\mu}\theta + K_{\mu}
\end{equation}
$K_{\mu}$ is not invariant under a color gauge transformation, but the RR field $\partial_{\mu}\theta$ will also transform under color on the D4 brane surface. This transformation cancels the 
variation of $K_{\mu}$ in (\ref{eq:RRfieldstrength}). This is the same basic idea as the Schwinger model Eq. (\ref{eq:KSdipole}). The connection between the color gauge field and the
bulk RR field can be described in the framework of anomaly inflow on the intersection of the D6 and D4 branes \cite{Green96}.

The pole cancellation enforced by gauge invariance requires that $\partial_{\mu}\theta$ should couple to the $q^2=0$ RR pole with the same strength as the CS current.
i.e. proportional to $\chi_t$. RR exchange in the quark-antiquark annihilation (hairpin) diagram, Fig. 1, reduces to a local
$\eta'$ mass term with the correct form of the Witten-Veneziano relation $m_0^2\propto \chi_t/f_{\pi}^2$,
\begin{equation}
\int d^4x d^4y \;\partial^{\mu}\eta'(x)/f_{\pi} G_{\mu\nu}(x-y) \partial^{\nu}\eta'(y)/f_{\pi} =
\frac{\chi_t}{f_{\pi}^2}\int d^4x \eta'^2(x)\;\; .
\end{equation}
Thus, RR exchange in the s-channel ($q\bar{q}$ annihilation) of quark-antiquark scattering produces a flavor singlet $\eta'$ mass insertion.
This effects only the singlet meson and breaks the axial $U(1)$ symmetry but leaves $SU(N_f)\times SU(N_f)$ chiral symmetry unbroken.
We also expect a contact term from RR exchange in the t-channel of the $q\bar{q}$ amplitude. 
This is the same for singlet and nonsinglet mesons. Written as a local 4-quark interaction, it is a $U(N_f)\times U(N_f)$ invariant Nambu-Jona Lasinio 
interaction. Note that the different flavor structure of s- and t-channel RR exchange simply reflects the familiar separation between valence and hairpin quark diagrams
in a lattice gauge calculation of meson propagators. The singlet meson correlator is the sum of the valence and hairpin diagrams, while the nonsinglet
correlator is given by the valence diagram alone. In the chiral limit, both the valence and hairpin diagrams have massless poles, which exactly cancel in the singlet channel
giving a massive $\eta'$. This cancellation is a direct manifestation of the Kogut-Susskind dipole cancellation, with the valence propagator representing the
Goldstone pole and the hairpin diagram representing the Chern-Simons correlator. 

To summarize, large-$N_c$ \cite{Witten_largeN} and holographic arguments \cite{Witten98}, as well as Monte Carlo evidence \cite{Horvath03,Ilgenfritz}, 
indicate that the topological structure of the QCD vacuum is dominated
by codimension one membranes, which play the role of domain walls separating $k$-vacua with $k$ units of Ramond-Ramond flux. This flux is described
by the local value of $\theta=2\pi k$, given by the value of the Wilson line of the RR field around $S_1$. 
These domain walls consist of singular excitations of the Chern-Simons tensor of the color gauge field with
support on 3-dimensional surfaces. The low energy chiral effects of these domain walls can be described in terms of Ramond-Ramond exchange and the
associated $q^2=0$ pole in the Chern-Simons correlator. These effects include phenomenologically important contact terms in QCD amplitudes. In the gauge field 
topological charge correlator the contact term is required for nonzero susceptibility. RR exchange in the $q\bar{q}$ scattering amplitude produces
the $\eta'$ mass insertion in the s-channel, while t-channel exchange gives a $U(N_f)\times U(N_f)$ invariant Nambu-Jona Lasinio 4-quark vertex which
could provide the pairing mechanism for spontaneous breaking of $SU(N_f)\times SU(N_f)$ chiral symmetry.

This work was supported by the Department of Energy under grant DE-FG02-97ER41027.

\end{document}